\documentclass[twocolumn,preprintnumbers,amsmath,amssymb,superscriptaddress,showpacs,prl]{revtex4}

\usepackage{graphicx}
\usepackage{dcolumn}
\usepackage{bm}

\begin{document}

\title{Emergence of a Dynamic Super-Structural Order Integrating Antiferroelectric and Antiferrodistortive Competing Instabilities in EuTiO$_3$}

\author{Jong-Woo Kim}

\email{jwkim@aps.anl.gov}

\affiliation{X-ray Science Division, Argonne National Laboratory, Argonne, IL, 60439, USA}

\author{Paul Thompson}
\affiliation{University of Liverpool, Dept. of Physics, Liverpool, L69 3BX, United Kingdom \& XMaS, European Synchrotron Radiation Facility, Grenoble, France}

\author{Simon Brown}
\affiliation{University of Liverpool, Dept. of Physics, Liverpool, L69 3BX, United Kingdom \& XMaS, European Synchrotron Radiation Facility, Grenoble, France}
\author{Peter S. Normile}
\affiliation{Universidad de Castilla-la Mancha, Spain}
\author{John A. Schlueter}
\affiliation{Materials Science Division, Argonne National Laboratory, Argonne, IL, 60439, USA}

\author{Andrey Shkabko}
\affiliation{Empa, Swiss Federal Laboratories for Materials Science and Technology, Ueberlandstrasse 129, 8600 Duebendorf, Switzerland}
\author{Anke Weidenkaff}
\affiliation{Empa, Swiss Federal Laboratories for Materials Science and Technology, Ueberlandstrasse 129, 8600 Duebendorf, Switzerland}

\author{Philip J. Ryan}
\email{pryan@aps.anl.gov}
\affiliation{X-ray Science Division, Argonne National Laboratory, Argonne, IL, 60439, USA}

\date{\today}

\begin{abstract}
Microscopic structural instabilities of EuTiO$_3$ single crystal were investigated by synchrotron x-ray diffraction. Antiferrodistortive (AFD) oxygen octahedral rotational order was observed alongside Ti derived antiferroelectric (AFE) distortions. The competition between the two instabilities is reconciled through a cooperatively modulated structure allowing both to coexist. The combination of electric and magnetic fields on the modulated AFD order shows that the origin of the large magnetoelectric coupling is based upon the dynamic equilibrium between the AFD - antiferromagnetic interactions versus the electric polarization - ferromagnetic interactions.
\end{abstract}


\maketitle

The magnetoelectric (ME) effect between electric and magnetic polarization is an interesting subject both in the fields of fundamental physics and materials science\cite{multiferro1}. Control of magnetic moment with electric field or electric polarization with magnetic field through the ME effect can drive new opportunities to develop future applications of low power field sensors, multi-state data storage and spintronic devices\cite{multiferro2}. However, typically this phenomenon is weak relegating device application unlikely\cite{multiferro1}.

The substantial change of the dielectric constant under an applied magnetic field observed in the EuTiO$_3$ system suggests a formidable coupling effect\cite{eto_dielec}. The rare-earth tetravalent titanate EuTiO$_3$ is one of the $A$TiO$_3$ perovskite members which presents quantum paraelectricity and G-type antiferromagnetic order of Eu moments below 5.3 K\cite{eto_mag_str}. Theoretical and experimental work predicted and confirmed that strained EuTiO$_3$, in thin film form exhibits ferromagnetic spin alignment as well as spontaneous electric polarization through spin-lattice coupling becoming a strong ferroelectric ferromagnet\cite{eto_theory,eto_film}. Recently, evidence of a cubic to tetragonal structural transition was reported, driven supposedly by TiO$_6$ octahedra rotations, analogous to SrTiO$_3$\cite{heat_cap,eto_xrd}. It is well established that AFD octahedral order competes directly with the electric polarization in tetravalent titanate perovskite systems\cite{afd_fe_inter}. These competing instabilities tend to suppress each other so that generally one prevails and determines the ground state structure. However, the ground state can be modified by external conditions, for example epitaxial strain or electric field taking advantage of a competitively balanced state\cite{strain}. A clear illustration of this effect was demonstrated by a series of artificial superlattice structures comprised of ferroelectric and paraelectric perovskite oxide components were designed to engineer high dielectric constants by tuning the competition between these instabilities\cite{superlattice1,superlattice2,superlattice3,superlattice4}.

In this work, we found that the EuTiO$_3$ system naturally forms a superlattice structure reconciling these competitive instabilities. Furthermore, since the magnetism is strongly coupled with the electric polarization as well as the oxygen octahedral rotations in this system\cite{unpub}, identifying the microscopic structure along with the combined response to electric and magnetic fields is essential to understand the ME coupling mechanism. We present x-ray diffraction data on a single crystal EuTiO$_3$ providing direct proof of AFD TiO$_6$ octahedral rotations and discuss their role within the ME coupling mechanism based on the interplay between AFD order and magnetic interactions.

Single crystals of EuTiO$_3$ were grown using a floating-zone furnace equipped with four focused halogen lamps and flowing mixture of 5 \% H$_2$ in Ar. X-ray diffraction measurements were performed on the 6ID-B beamline at the Advanced Photon Source and the XMaS beamline at European Synchrotron Radiation Facility. The sample was mounted on the cold finger of a Joule-Thomson stage in a closed cycle helium displex refrigerator modified to provide \emph{in-situ} high electric field application. The incident x-ray energy was tuned to 16.2 keV for the structural measurement and 7.612 keV, Eu L$_{II}$ edge was used for the x-ray resonant magnetic scattering (XRMS) measurement. The incident x-ray is linearly polarized perpendicular to the scattering plane ($\sigma$ polarization). The resonant magnetic scattering arising from electric dipole (E1) transitions from the $2p$-to-$5d$ states, rotates the polarization resulting in $\pi$ polarized photons (parallel to the scattering plane). Polarization analysis was achieved by using pyrolytic graphite (PG) (0~0~6) reflection to select $\pi$ -polarized magnetic scattering and suppress the background from charge scattering ($\sigma$ polarized).  The sample was thinned down to 400 microns and coated with Au electrodes in order to apply electric field along the [1~1~0] direction.

\begin{figure}
\begin{center}
\includegraphics[clip, angle=0, width=0.5\textwidth]{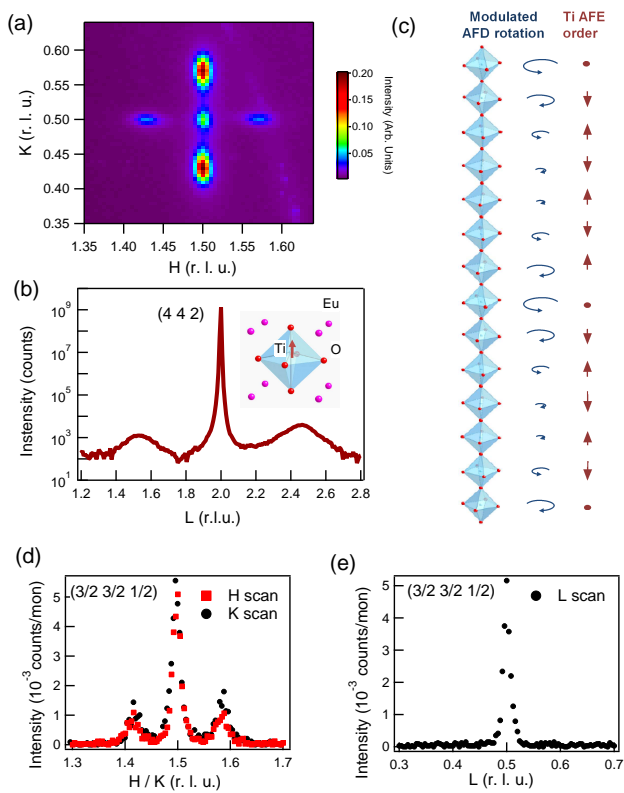}
\caption{\label{fig:fig1} (Color online) (a) Reciprocal space map in H and K of the  antiferrodistortive octahedral order reflection at (3/2~1/2~5/2) presenting the modulation satellites. (b) L scan through the (4~4~2) Bragg reflection and an atomic model of the titanium displacement. Broad width intensity is obserbed at (4~4~2$\pm$1/2) corresponding to the local AFE order. (c) An atomic model of the modulated AFD order and the corresponding AFE order. The arrows show the possible local electric polarization order but the actual titanium shift direction is not determined from this measurement.(d) H and K scans around (3/2~3/2~1/2). In both scans, the m-AFD reflections are allowed due to H,K$\neq$L (e) L scan around (3/2~3/2~1/2). The m-AFD reflection is forbidden as H=K.}
\end{center}
\end{figure}

Half integer Bragg peaks were observed, arising from AFD octahedral tilting (central reflection in Fig.~\ref{fig:fig1} (a)). In principle, the symmetry of the octahedral rotation patterns can be identified by the reflection condition of these peaks. It is assumed the system undergoes a structural phase transition from cubic to tetragonal symmetry, $a^0a^0c^-$ in Glazer\cite{glazer,eto_octa_theory} notation similar to SrTiO$_3$ and any a, b, or c axis of the cubic unit cell can become the tetragonal c-axis\cite{eto_xrd}. However, it is problematic to distinguish the single crystal c-axis due to small lattice changes and broad mosaicity in order to apply reflection conditions to identify the symmetry correctly.

Additional reflections illustrated in Fig. 1(a) are found flanking the half order reflections along H, K and L directions. The wave vector of these peaks is (1/2~1/2~q), (1/2~q~1/2) and (q~1/2~1/2) where q = $\sim$0.43 indicates an incommensurate superstructure periodicity $\sim$14 unit cells. The reflection condition states that if two of the H, K and L are equal, then the satellite reflection is forbidden (Fig.~\ref{fig:fig1} (d) and (e)). This indicates that the satellite peaks result from the modulated AFD (m-AFD) octahedral rotation and that the rotation axis is along the q direction. While oxygen atoms lying on the rotation axis do not change their position by rotation, the other oxygen atoms move from the face center of the perovskite unit cell. Since oxygen atoms share the position with the next unit cell the wave vector components perpendicular to the rotation axis are constrained to be 1/2.

Concurrently, weak intensities of (0~0~1/2) order also emerge around the (4~4~2) reflection corresponding to an AFE order due to Ti displacements at base temperature, shown in Fig.~\ref{fig:fig1} (b). The associated correlation length of $\sim$6 unit cells is roughly half the length of the m-AFD rotational order. The m-AFD generates regions of both larger and smaller rotation angles of TiO$_6$ octahedra. The structure model in Fig.~\ref{fig:fig1}  (c) illustrates how the short range AFE periodicity forms where the AFD rotations are near a minimum. In order for both AFD and AFE orders to coexist, the competition between them is reconciled through the formation of the super structure where both instabilities are alternatively interwoven.

\begin{figure}
\begin{center}
\includegraphics[clip, angle=0, width=0.5\textwidth]{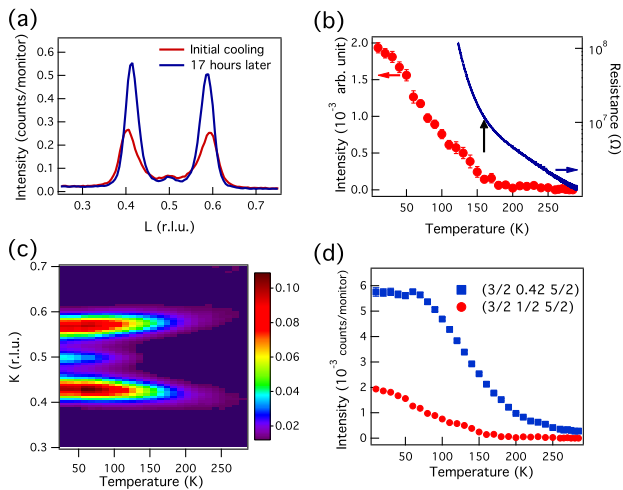}
\caption{\label{fig:fig2} (Color online) (a) The L scans around (7/2~5/2~1/2) right after cooling and 17 hours later. The intensity of the modulated AFD order increases slowly over long time. (b) The temperature dependence of the AFD intensity at (3/2~1/2~5/2) and the resistance of the sample. The response of resistance variation by temperature changes around 160 K where the AFD order occurs. (c) Temperature dependence of the K scans in color scale. (d) The temperature dependence of integrated intensities of normal and modulated AFD order reflections. The normal AFD order disappears around 160 K while the modulated AFD order persists up to 285 K.}
\end{center}
\end{figure}

Additionally, after cooling the superlattice modulation continues to develop slowly over time. Figure~\ref{fig:fig2} (a) shows the evolution of the m-AFD reflection by comparing immediately upon cooling and after 17 hours. The m-AFD peaks increase in intensity, shift in position and sharpen over long time periods while the simple AFD intensity remains unchanged. This implies that there is a large relaxation time constant for m-AFD structure to form indicating the mediation is a dynamic process with continuing fluctuations between the AFD and AFE order. In contrast, the simple AFD order is static. The correlation lengths of the m-AFD structure are within the nanometer regime, $\sim$11 nm (28 unit cells) and $\sim$22 nm (56 unit cells)  parallel and perpendicular to the octahedral rotation axis direction respectively.

\begin{figure}
\begin{center}
\includegraphics[clip, width=0.5\textwidth]{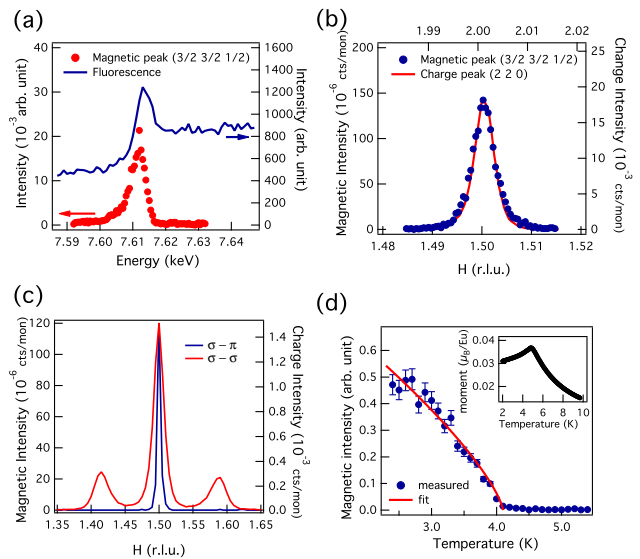}
\caption{\label{fig:fig3} (Color online)(a) Energy scans of the magnetic reflection (3/2 3/2 1/2) in $\sigma-\pi$ geometry and fluorescence. The resonance enhancement is shown at the Eu L$_{II}$ absorption edge. (b) H scans around magnetic reflection (3/2~3/2~1/2) and charge Bragg peak (2~2~0). The width of the magnetic peak is comparable with the width of the charge peak. (c) H scans around (3/2~3/2~1/2) reflection in  $\sigma-\sigma$  and $\sigma-\pi$  polarization geometries. Only magnetic intensity from G-type magnetic order is shown in $\sigma-\pi$  geometry while the AFD order reflections are seen in $\sigma-\sigma$ geometries.  (d) Temperature dependence of the magnetic intensity and the critical exponent fitting curve. Inset shows the SQUID measurement of magnetization versus temperature curve with 100 Oe}
\end{center}
\end{figure}

The temperature dependence of the K scan across (3/2~1/2~5/2) is plotted in Fig.~\ref{fig:fig2} (c) and (d). The m-AFD peak intensity disappears around 285 K, which coincides with a transition found by heat capacity measurements\cite{heat_cap}, however the simple AFD reflection vanishes around $\sim$160 K. Additionally, the incommensurate periodicity contracts with increasing temperature accelerating as the AFD order dissipates. The resistivity also shows an transition at this temperature shown in figure Fig.~\ref{fig:fig2} (b), which indicates a band gap broadening with static symmetry reduction from octahedral rotations\cite{gap_open1}. In fact, controlling the gap by strain has been calculated in SrTiO$_3$, by changing the degree of oxygen rotation, the O $2p$ and Ti $3d$ states are more likely to mix and consequently repel each other, essentially driving the respective valence and conduction bands further apart\cite{gap_open2}.

The G-type antiferromagnetic structure was confirmed by x-ray resonant magnetic scattering. Figure~\ref{fig:fig3} (a) presents the resonant enhancement of the (3/2~3/2~1/2) magnetic reflection intensity at the Eu L$_{II}$ edge below T$_N$. Figure~\ref{fig:fig3} (b) shows that the width of the magnetic peak along H is comparable to the width of the normal structural Bragg peak indicating the correlation length of magnetic order is comparable to the size of the crystal grain. The clear difference between the AFM and m-AFD correlation lengths demonstrates how the m-AFD order is not associated with crystal quality(Fig.~\ref{fig:fig3} (c)). An estimated T$_N$ and critical exponent were extracted from the temperature dependent XRMS intensity in figure~\ref{fig:fig3} (d), as 4.1K and 0.373 respectively showing a 3D Heisenberg behavior. The transition temperature measured by XRMS is slightly lower than the SQUID measurement due to the x-ray beam heating effect.

\begin{figure}
\begin{center}
\includegraphics[clip, angle=0, width=0.35\textwidth]{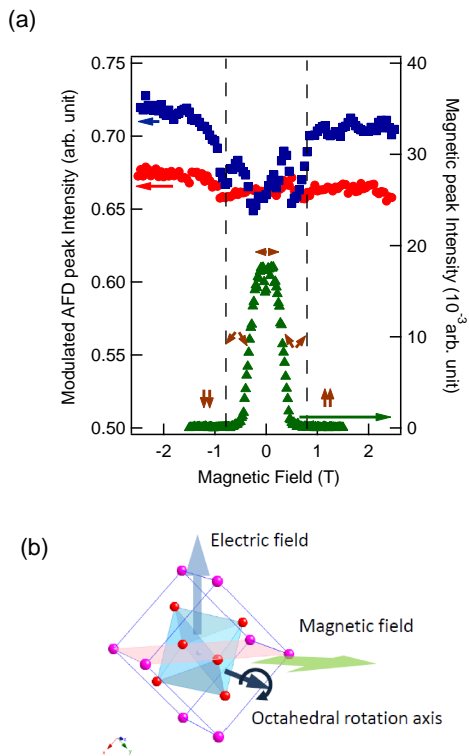}
\caption{\label{fig:fig4} (Color online) (a) The magnetic field effects on the modulated AFD order at (7/2~5/2~q) reflection and AFM order at (3/2~3/2~1/2) magnetic reflection. The red circle is the intensity of modulated AFD order without E-field application. The blue square is measured with E-field, 0.67 $\times$10$^5$ V/cm. About 10 \% of the intensity change is shown around the saturation field ($\sim$0.7 T) with E-field application. The green triangle is intensity of the AFM order and it decreases during the magnitude of magnetic field increases and disappear when the magnetic moments are saturated. The brown arrows show the response of the magnetic moment alignments by external magnetic field. (b) The configuration of the electric and magnetic field directions according to the crystallographic geometry}
\end{center}
\end{figure}

The magnetic intensity of AFM ordering as a function of applied B-field along the [1~1~1] direction is plotted in Figure~\ref{fig:fig4}. Canting of the magnetic moment occurs along the external magnetic field direction and becomes fully aligned to the field above 0.7 T. The saturation field is lower than the previous measurement on the powder sample (between 1 and 3 T) suggesting that the [1~1~1] direction is the magnetic easy axis. The magnetic response of the m-AFD (7/2~5/2~q) reflection was measured by sweeping the B-field along the [1~1~1] direction, with and without applied electric field along the [1~1~0] direction as illustrated in figure~\ref{fig:fig4} (b).

A small intensity change was observed at the magnetic field saturation points without E-field application. No significant change to the m-AFD reflection is measured with E-field alone and similarly, no effect is observed on the magnetic reflection intensity with E-field either (not shown). Additionally, the magnetic field required to saturate the system remains unchanged with E-field, implying that the maximum E-field $\sim$0.67$\times$10$^5$ V/cm may not be sufficient to alter the antiferromagnetic interaction\cite{unpub2}. However, a large change of the m-AFD intensity was found when both E and B fields are applied simultaneously. The measurement was made while the E-field was fixed at 0.67$\times$10$^5$ V/cm and the applied B-field swept from -2.5 T to 2.5 T.  A large increase in intensity, up to $\sim$10 \%, is observed through the saturation point, not seen with either E or B field application alone.

This establishes the central role the m-AFD order plays in the underlying mechanism of the magneto-dielectric coupling in this system. As was discussed above, the competition of the octahedral rotation and the electric polarization is accommodated by forming the m-AFD order. It is a dynamic equilibrium state with continuous and coupled fluctuations between the AFD and AFE instabilities. Hence, this delicate balance can be modified by external conditions more readily. It is known that the titanium shift is related to ferro- and antiferromagnetic spin alignment through the spin-lattice coupling\cite{eto_dielec}. In addition, recent calculations show that the octahedral rotations are indirectly linked to the AFM magnetic interaction energy again through the titanium position\cite{unpub}. The external electric and magnetic fields alter the dynamically coupled equilibrium state of the AFD and AFE instabilities. As a result, the system responds by shifting to a new equilibrium position and subsequently increases the population of m-AFD order.

In conclusion, we have revealed a novel dynamic microscopic superstructural response reconciling competing AFD and electric polar instabilities in EuTiO$_3$ single crystals, by employing synchrotron x-ray diffraction. Due to the competition between the AFD octahedral rotation and electric polarization, the local structure approaches a dynamic equilibrium state with a large time scale, resulting in a modulated AFD order. By forming this structure, the coexistence of the competing AFD and AFE structural instabilities becomes possible. The equilibrium can be tuned by external electric and magnetic field application, indicating that the m-AFD order is key to understanding the magnetoelectric nature of this system.

Work at Argonne and use of beamline 6-ID-B at the Advanced Photon Source at Argonne was supported by the U. S. Department of Energy, Office of Science, Office of Basic Energy Sciences under Contract No. DE-AC02-06CH11357. The EPSRC-funded XMaS beamline at the ESRF is directed by M.J. Cooper, C.A. Lucas and T.P.A. Hase. We are grateful to O. Bikondoa, D. Wermeille, and L. Bouchenoire for their invaluable assistance and to S. Beaufoy and J. Kervin for additional XMaS support. Funding for sample growth is provided by SNF NCCR MaNEP. P. J. Ryan acknowledges additional funding from the University of Liverpool and Aer Lingus plc.




\begin{thebibliography}{}
\bibitem{unpub2} The antiferromagnetic interaction is determined by the third nearest neighbor Eu ion mediated by a superexchange interaction via Eu-Ti-Eu. Displacing the Ti atom by external electric field should alter the balance of antiferromagnetic and ferromagnetic interaction preferentially suppressing the antiferromagnetic interaction.\cite{unpub}
\end{thebibliography}
\end{document}